\documentclass[proceedings]{elsarticle}
\usepackage[hmargin=3cm]{geometry}

\makeatletter
\def\ps@pprintTitle{%
	\let\@oddhead\@empty
	\let\@evenhead\@empty
	\def\@oddfoot{\centerline{\thepage}}%
	\let\@evenfoot\@oddfoot}

\usepackage{lineno,hyperref}
\usepackage{siunitx}
\usepackage{caption}
\usepackage{subcaption}
\usepackage{xcolor}
\begin{document}

\begin{frontmatter}

\title{The Mu3e experiment: Toward the construction of an HV-MAPS vertex detector}


\author[HD]{Thomas Rudzki\corref{correspondingauthor}}

\cortext[correspondingauthor]{Corresponding author} 
\ead{rudzki@physi.uni-heidelberg.de}

\author[HD]{Heiko Augustin}
\author[FH]{Marin Deflorin}
\author[HD]{Sebastian Dittmeier}
\author[HD]{Florian Frauen}
\author[HD]{David M. Immig}
\author[HD]{Dohun Kim}
\author[PSI]{Frank Meier Aeschbacher}
\author[HD]{Annie Meneses Gonz\'{a}lez}
\author[HD]{Marius Menzel}
\author[KA]{Ivan Peri\'{c}}
\author[HD]{Sebastian Preu\ss}
\author[HD]{Andr\'{e} Sch\"oning}
\author[HD]{Luigi Vigani}
\author[HD,KA]{Alena Weber}
\author[HD]{\\Benjamin Weinl\"ader}

\address[PSI]{Paul Scherrer Institut, 5232 Villigen, Switzerland} 
\address[HD]{Universit\"at Heidelberg, Physikalisches Institut, 69120 Heidelberg, Germany}
\address[KA]{Karlsruhe Institute of Technology, Institute of Data Processing and Electronics, 76021 Karlsruhe, Germany}
\address[FH]{Fachhochschule Nordwestschweiz, Institut f\"ur Thermo- und Fluid-Engineering, 5210 Windisch, Switzerland}

\begin{abstract}

The Mu3e experiment searches for the lepton flavor violating decay $\mu^+~\rightarrow~e^+~e^+~e^-$ with an ultimate aimed sensitivity of 1 event in $10^{16}$ decays. 
This goal can only be achieved by reducing the material budget per tracking layer to $X/X_0\approx \SI{0.1}{\percent}$.
High-Voltage Monolithic Active Pixel Sensors (HV-MAPS) which are thinned to \SI{50}{\micro\meter} serve as sensors. 
Gaseous helium is chosen as coolant.

Results of recent studies related to the sensor prototypes, the helium cooling, and module prototyping are presented.
The recent chip submission MuPix10 has proven its functionality regarding efficiency and time resolution.
The helium cooling system for the inner tracker could be verified using a full-scale prototype.
A complete prototype equipped with MuPix10 chips will be tested inside the Mu3e magnet in summer 2021.

\end{abstract}

\begin{keyword}
	pixel detector \sep Mu3e \sep MuPix  \sep HV-MAPS \sep HV-CMOS \sep cooling
\end{keyword}

\end{frontmatter}


\section{Introduction}

The Mu3e experiment searches for the charged lepton flavor violating decay of $\mu~\rightarrow~e^+~e^-~e^-$~\cite{mu3e:research_proposal}.
It aims to exploit branching ratios up to $10^{-16}$.
The experiment will be located at Paul Scherrer Institut, Switzerland.
The current limit on the branching ratio of this process was set to be $<10^{-12}$ by SINDRUM in 1988~\cite{sindrum}.

Such an increase in sensitivity is only possible by using modern detectors as pixel sensors and drastically reducing their material budget since multiple-Coulomb scattering is the most dominant background source.
Specific high-voltage monolithic active pixel sensors (HV-MAPS) have been developed for this experiment~\cite{hvmaps_ivan}.
It is possible to thin the chips down to \SI{50}{\micro\meter}.
Tight constraints on the cooling, electronics and support structure have to be tackled to minimize the material budget.

In the following, recent results in sensor development will be presented.
The construction and testing of thermal-mechanical mock-ups of the Mu3e vertex detector with a strong emphasis on the gaseous helium cooling and a first functional detector prototype are shown.
The vertex detector forms the central part of the experiment with its two pixel layers (Figure~\ref{fig:mu3e_schematic}).
The main goal of prototyping is to achieve production readiness for the final detector.
An overview on the developed custom tooling and working steps will be given.

\section{The Mu3e pixel sensors}

The Mu3e tracking detectors use HV-MAPS.
The ASIC called MuPix is produced in a commercially available \SI{180}{\nano\meter} HV-CMOS process.
Charges induced by traversing particles are collected via drift at the diode which enables good time resolution of a few nanoseconds~\cite{hvmaps_ivan}.
The readout ASIC is fully integrated on chip.
The sensors are thinned to \SI{50}{\micro\meter}.
Thus, this technique can significantly reduce the material budget compared to conventional hybrid pixel sensors. 
The MuPix prototype series proved full functionality ranging from high efficiency of $>\SI{99}{\percent}$ to time resolutions of $6-\SI{8}{\nano\second}$~\cite{proceedings_andre}.

The Mu3e requirements~\cite{tdr} for the pixel sensors are:

\begin{itemize}
	\setlength\itemsep{-0.5em}
	\item Efficiency $>\SI{99}{\percent}$
	\item Noise rate $<\SI{20}{\hertz/pixel}$
	\item Time resolution $<\SI{20}{\nano\second}$
	\item $\sim\SI{2x2}{\centi\meter}$ active pixel matrix
	\item \SI{80x80}{\micro\meter} pixel size
	\item Thickness of \SI{50}{\micro\meter}
	\item Maximum power dissipation of \SI{350}{\milli\watt/\centi\meter^2}
\end{itemize}

In addition, space constraints on the two-layered high-density interconnects which provide all electrical connections limit the number of input voltages to one low voltage (LV) and one high voltage (HV) line.

\begin{figure*}[bht]
	\centering
	\begin{subfigure}[c]{.50\textwidth}
		\includegraphics[width=\textwidth]{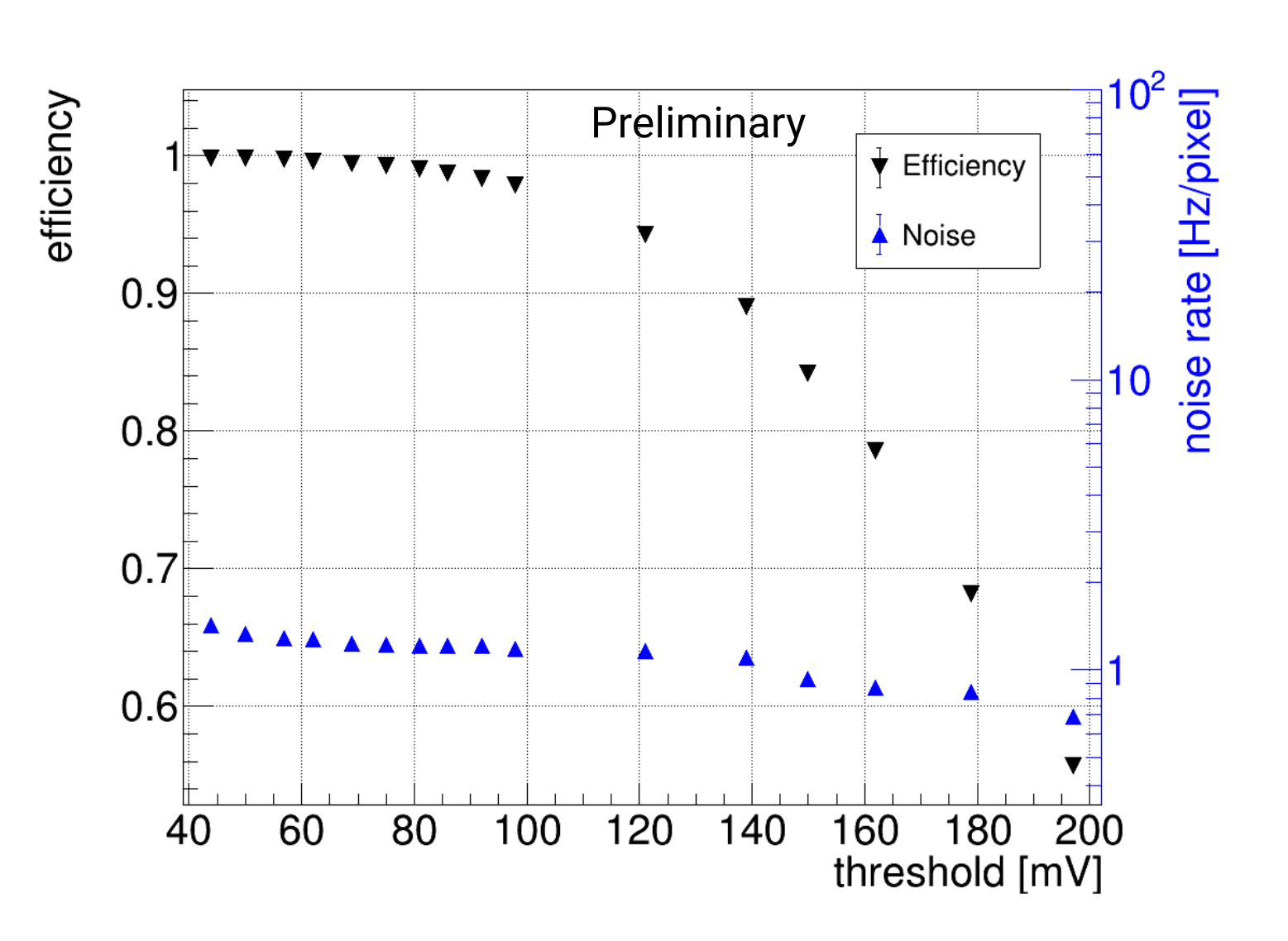}
		\caption{MuPix10 single hit efficiencies and noise rate as function of the discriminator threshold~\cite{bttb9_david}.}
		\label{fig:mpx_scan}
	\end{subfigure}
	\hspace{0.5em}
	\begin{subfigure}[c]{.35\textwidth}
		\bigskip
		\includegraphics[width=\textwidth]{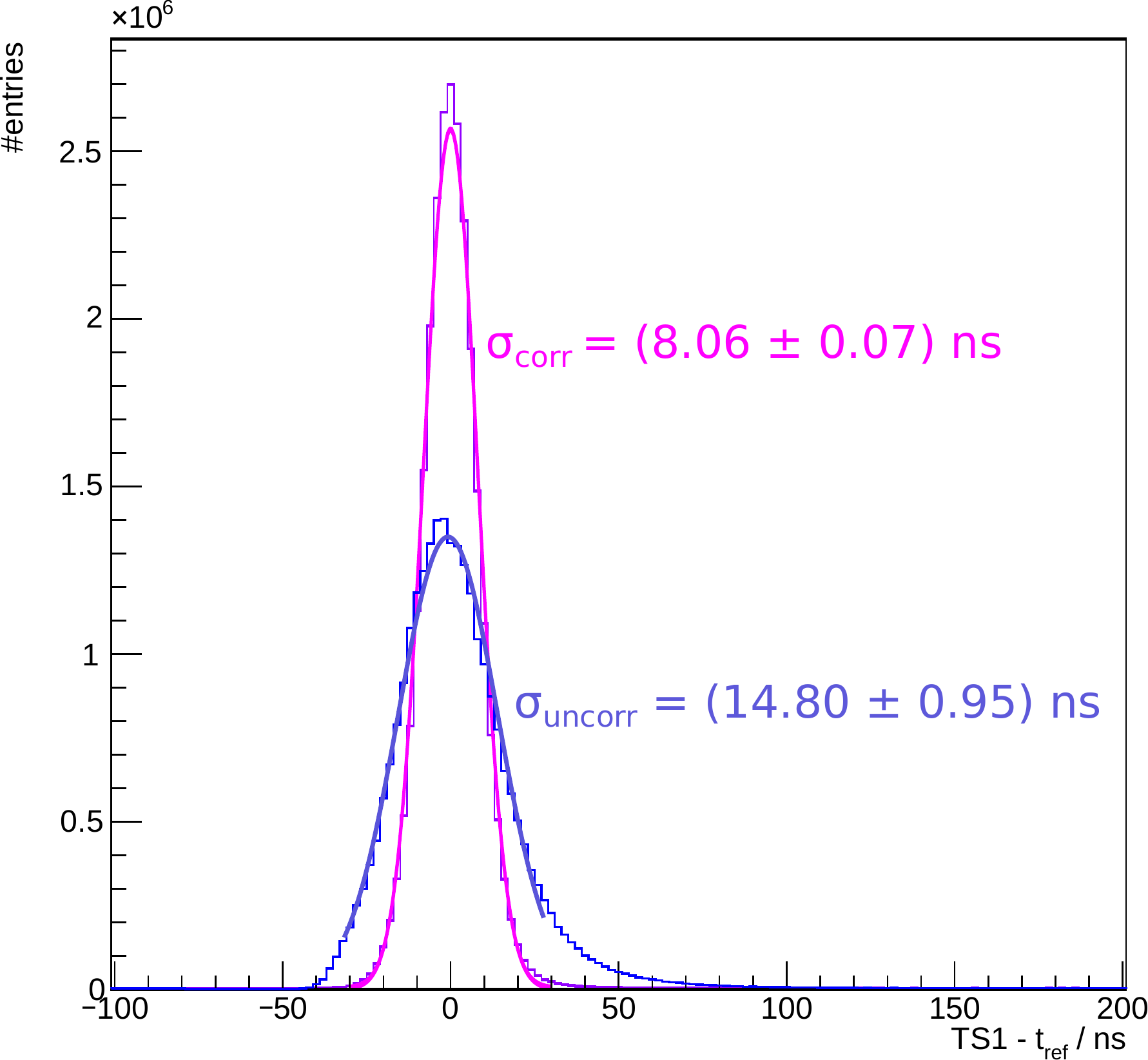}
		\caption{MuPix10 time resolution; blue colored raw data, pink colored including time-walk and delay corrections~\cite{ba_frauen}.}
		\label{fig:mpx_time}
	\end{subfigure}
	\caption{MuPix10 efficiency and noise rate results obtained at a \SI{5}{\giga\electronvolt} electron beam at DESY, time resolution results from lab measurements.}
	\label{fig:mpx_desy}
\end{figure*}

\paragraph{Sensor characterization}
The most recent prototype is the MuPix10 sensor~\cite{mupix10} which is the first full-scale prototype of the MuPix series.
The sensors (\SI{100}{\micro\meter} thickness) were characterized in a test beam campaign at DESY and in lab measurements in Heidelberg.
Digital and analog supply lines were shorted on the test PCBs of MuPix10.

Figure~\ref{fig:mpx_scan} shows that MuPix10 reaches efficiencies of $>\SI{99}{\percent}$ for a noise rate below \SI{2}{\hertz/pixel} in test beam using \SI{5}{\giga\electronvolt} electrons.
Lab studies achieved a time resolution of \SI{14.80}{\nano\second} (raw) and \SI{8.06}{\nano\second} including time walk and delay corrections for the point-to-point transmission line from the pixel to the periphery~\cite{ba_frauen} (Figure~\ref{fig:mpx_time}).
The test beam results were obtained with a power dissipation of slightly less than \SI{200}{\milli\watt/\centi\meter^2}. 


The tested \SI{100}{\micro\meter} thin sensors fulfill the Mu3e requirements. 
Studies with sensors thinned to \SI{50}{\micro\meter} are ongoing.

\section{The Mu3e pixel detectors}

\begin{figure}[h!]
	\centering
	\includegraphics[width=\textwidth]{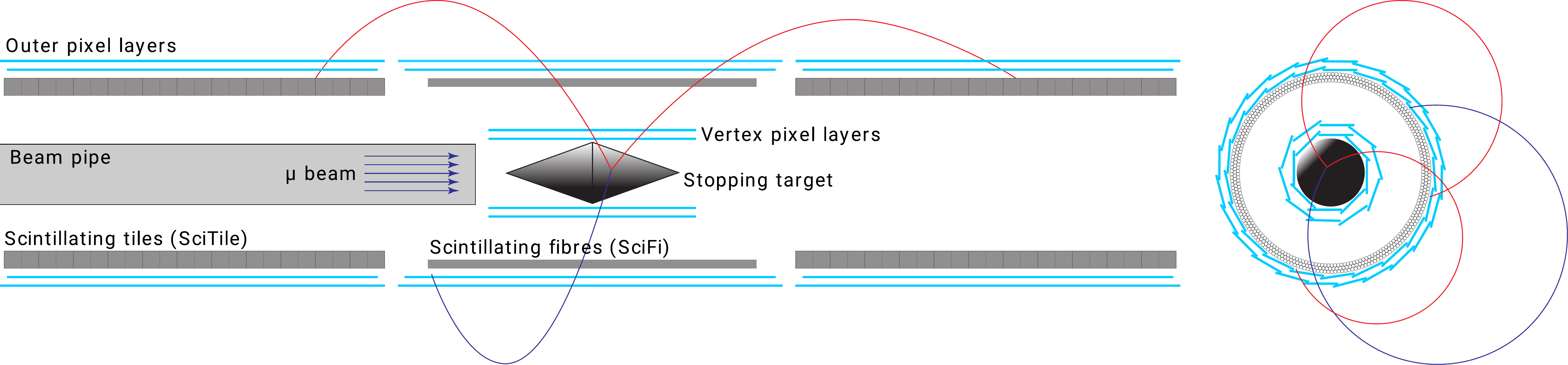}
	\caption{Schematic view on the Mu3e experiment, longitudinal and transverse view with indicated tracks for a $\mu~\rightarrow~e~e~e$ event \cite{proceedings_frank}.}
	\label{fig:mu3e_schematic}
\end{figure}

\begin{figure}[tb!]
	\centering
	\includegraphics[width=.7\textwidth]{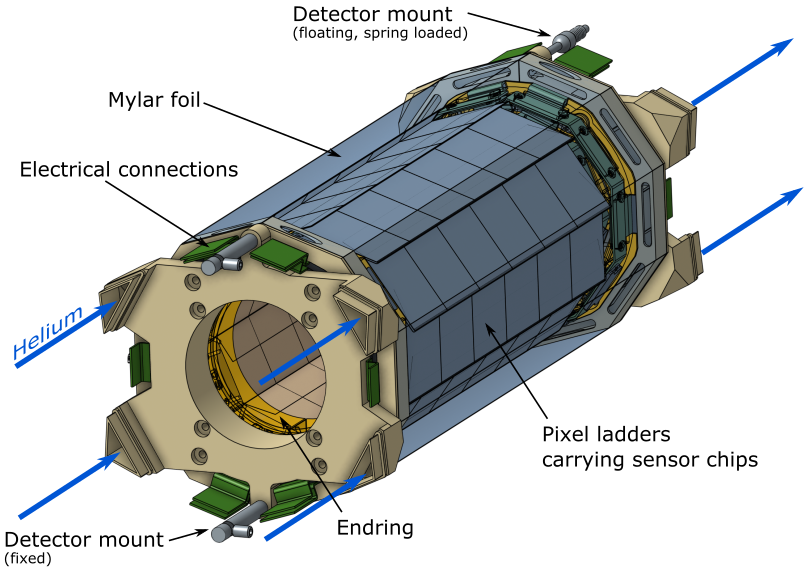}
	\caption{Schematic view of the Mu3e vertex detector including the mechanical support and helium distribution rings~\cite{tdr}.}
	\label{fig:vertex_detector}
\end{figure}

The Mu3e pixel detector is divided in the inner layers forming the vertex detector and the outer layers (Figure~\ref{fig:mu3e_schematic}).
The vertex detector (Figure~\ref{fig:vertex_detector}) consists of two pixel layers surrounding the muon target.
Its layers are made of 8/10 ladders which carry six sensors each.

In turn, the outer layers form three stations with two layers each. 
The central one surrounds the vertex detector and the target while two additional stations are located upstream and downstream.
The layers are made of 24/28 ladders which carry 17/18 sensors.

\begin{figure*}[th!]
	\centering
	\begin{subfigure}[b]{.7\textwidth}
		\includegraphics[width=\textwidth]{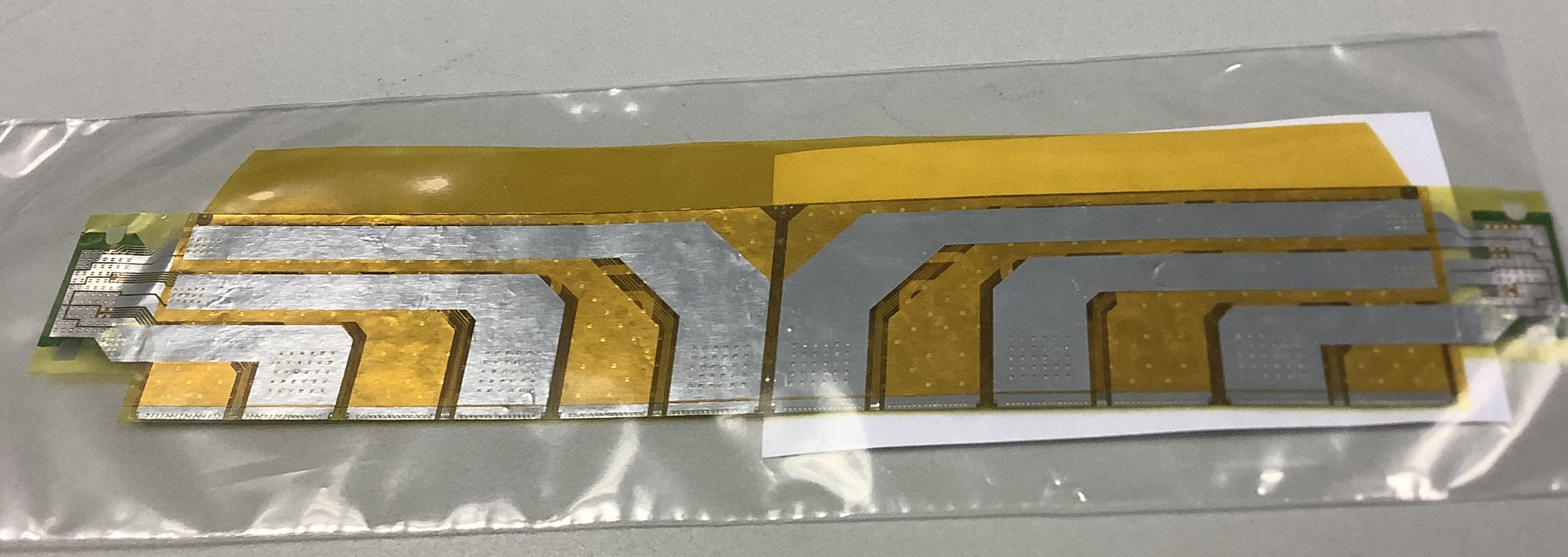}
		\caption{High-density interconnect (HDI) for silicon heater chips produced at LTU, Ukraine~\cite{LTU}.}
		\label{fig:hdi}
	\end{subfigure}
	\hspace{0.5em}
	\begin{subfigure}[b]{.2\textwidth}
		\bigskip
		\includegraphics[width=\textwidth]{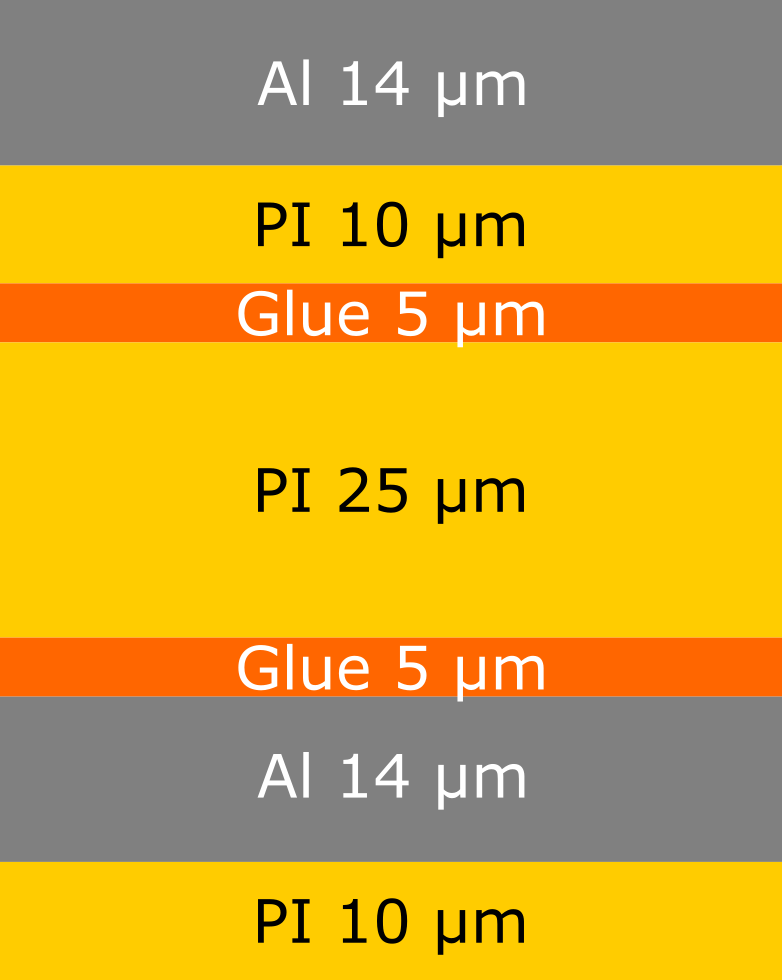}
		\caption{Layer stack of an HDI~\cite{tdr}.}
		\label{fig:hdi_stack}
	\end{subfigure}
	\caption{}
	\label{fig:silicon_heater_measurement}
\end{figure*}

A ladder consists of a high-density interconnect (HDI, Figure~\ref{fig:hdi}) and the MuPix sensors only. 
An~HDI is made of two aluminized polyimide layers and a spacer polyimide layer (Figure~\ref{fig:hdi_stack}) with an overall thickness of $\sim\SI{80}{\micro\meter}$.
It provides all electrical connections as HV, LV and signal lines and serves as only support structure.
The electrical connections to the chip are established via \textit{single-point Tape Automated Bonding} (spTAB).
This results in an overall radiation length of only $X/X_0 \approx \SI{0.115}{\percent}$ per layer.

\section{Helium cooling system}
\label{sec:helium}

The pixel detectors are cooled by a helium gas flow.
This keeps the material budget of the experiment low.
The maximum allowed temperature for the ladders is \SI{70}{\degreeCelsius} which is the glass transition temperature of the adhesives used for the construction of the detector.

In the vertex detector a helium flow between the two layers and another flow that is confined by layer~2 and a mylar cylinder with a total mass flow of \SI{2}{\gram/\second} is realized.
The helium flow is guided by ducts along the beam pipe to the detectors~\cite{tdr}.
On the faces of the barrel-shaped detector, gas distribution rings are integrated in the support structure (Figure~\ref{fig:vertex_detector}).

\begin{figure}[tb!]
	\centering
	\includegraphics[width=0.75\textwidth]{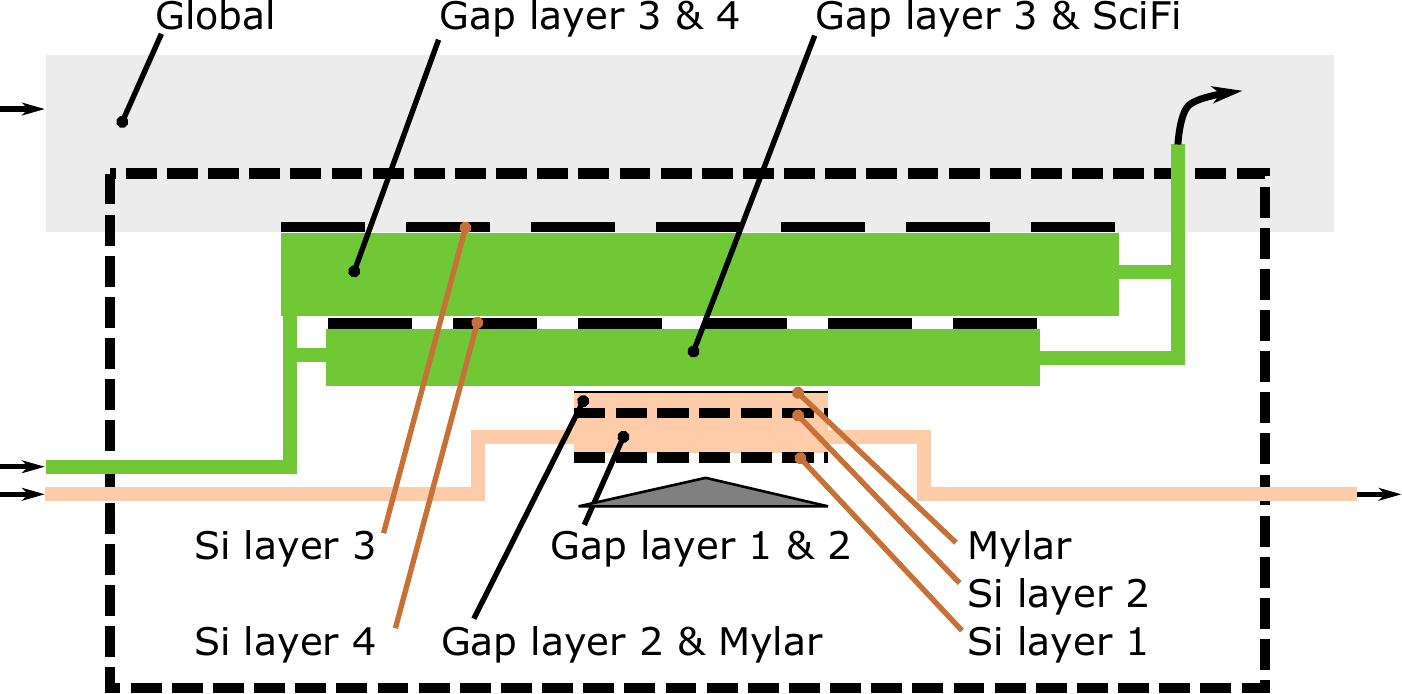}
	\caption{Sketch of the helium flows for cooling the central detector station. Pink is the gas flow for the vertex detector, green is the gas flow of the outer layers, light gray indicates an additional global flow, dark gray shows the muon target.}
	\label{fig:he_cooling}
\end{figure}

\begin{figure}[tb!]
	\centering
	\includegraphics[width=0.9\textwidth]{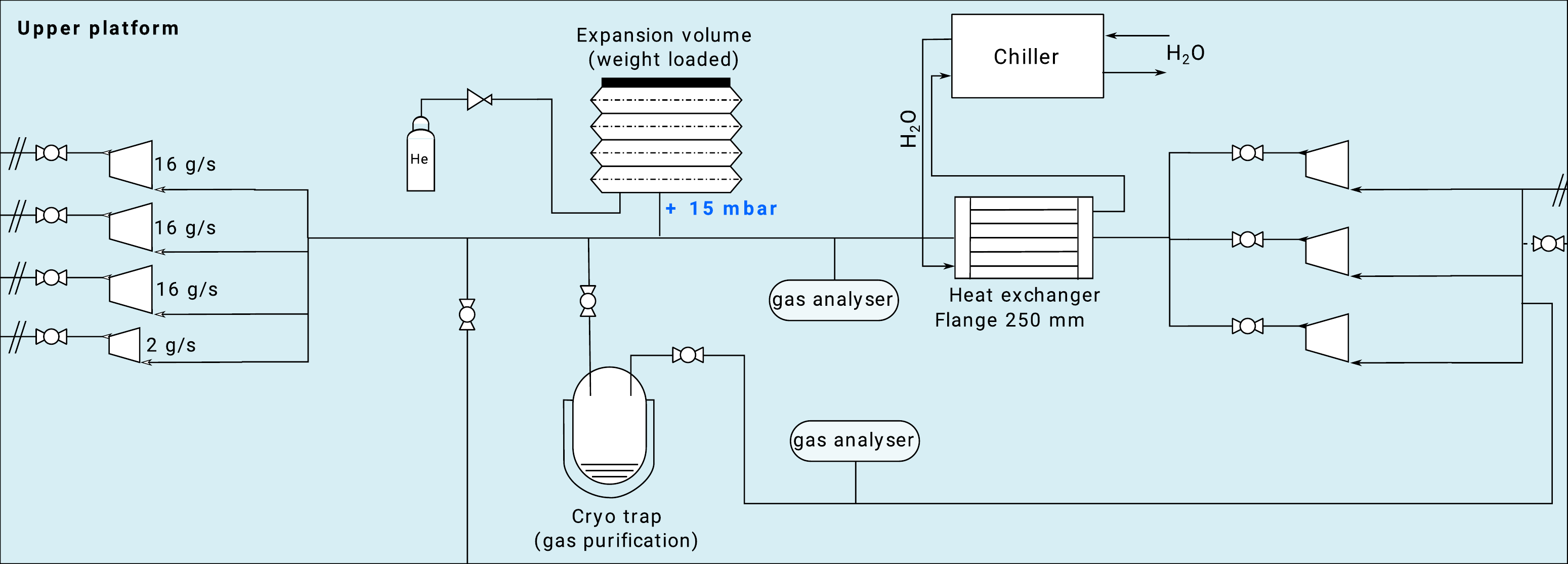}
	\caption{Sketch of the helium infrastructure outside the Mu3e magnet. The trapezoidal structures show the turbo compressors.}
	\label{fig:he_system}
\end{figure}

The infrastructure to provide the helium flows is located outside the Mu3e magnet.
It consists of a heat exchanger with a chiller, an expansion volume to ensure constant pressure, a cryo trap for gas purification and a set of turbo compressors on the inlet and outlet (Figure~\ref{fig:he_system}).
The main challenge is to circulate helium with a high mass flow and a low pressure difference between inlet and outlet.
This is realized by miniature turbo compressors.
The inlet temperature of the helium is \SI{0}{\degreeCelsius}~to~\SI{4}{\degreeCelsius}.

A prototype helium cooling system that is able to provide a mass flow of \SI{2}{\gram/\second} making use of one turbo compressor was realized at Fachhochschule Nordwestschweiz (FHNW) in Windisch, Switzerland.
Results obtained with this cooling setup are described in section~\ref{sec:prototypes}.

\section{Vertex detector construction}
\label{sec:construction}

The Mu3e vertex detector will be mounted and tested at the Physics Institute in Heidelberg.
Due to the small number of ladders and sensors, the whole construction procedure will be done manually.
Custom tooling was designed to enable simple and safe handling of the detector components.
To validate the manufacturing procedures, prototypes have been produced.

\begin{figure*}[bt]
	\centering
	\begin{subfigure}[b]{.58\textwidth}
		\includegraphics[width=\textwidth]{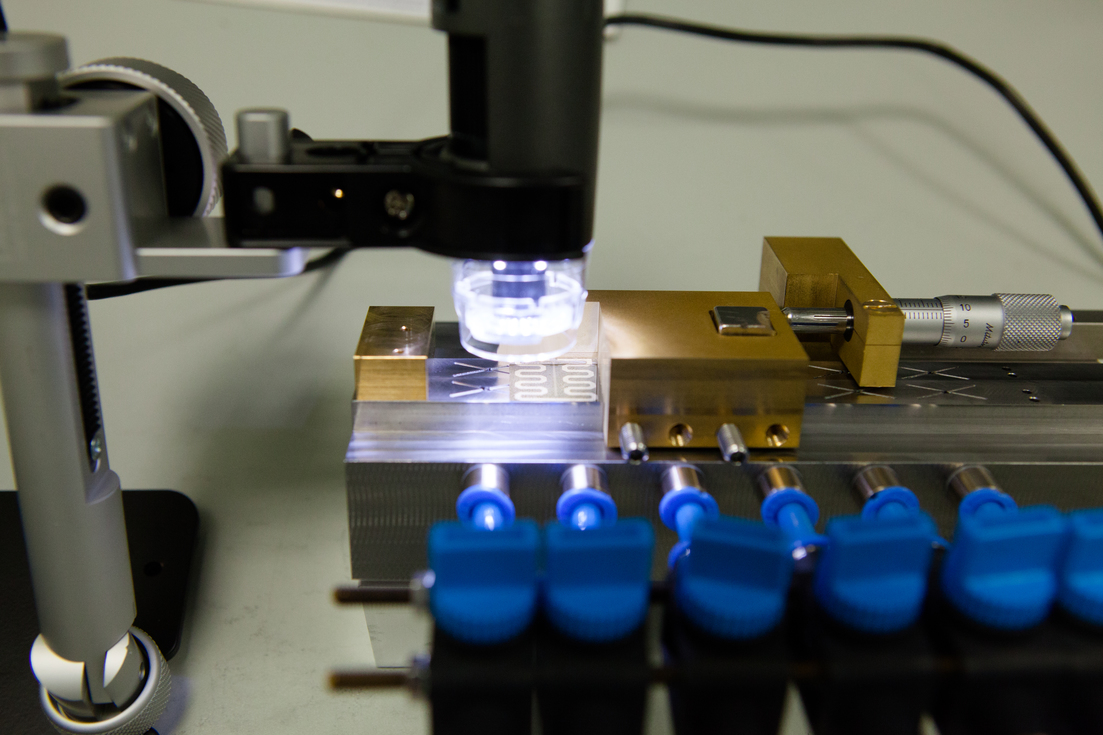}
		\caption{Sliding block to position sensor with micrometer screw, position monitored by digital microscope~\cite{tdr}.}
		\label{fig:chip_placement}
	\end{subfigure}
	\hspace{0.5em}
	\begin{subfigure}[b]{.37\textwidth}
		\bigskip
		\includegraphics[width=\textwidth]{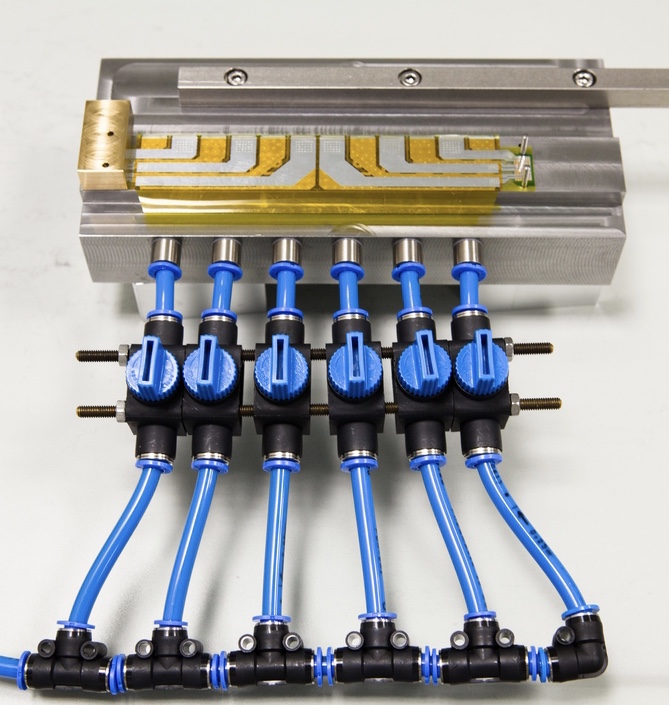}
		\caption{Ladder after gluing the HDI to six chips on the vacuum chuck~\cite{tdr}.}
		\label{fig:ladder_assembly}
	\end{subfigure}
	\caption{Mounting tool for chip placement and ladder assembly}
	\label{fig:tooling1}
\end{figure*}

\paragraph{Ladder construction}
In a first step, the six chips are positioned relative to each other.
The most stringent constraint is the placement precision along the beam axis of $\sigma_z = \SI{5}{\micro\meter}$.
The chips are placed one by one with a sliding block that is moved by a precise micrometer screw (Figure~\ref{fig:chip_placement}).
The chip position is monitored with a digital microscope\footnote{Dino-Lite AM4515T8-EDGE} with an optical resolution of \SI{1.5}{\micro\meter}.
After reaching the position the sensors are fixated by vacuum.
On each side, flexible printed circuit boards (flexprints) to connect the ladder to the outside are placed using reference pins.

Next, epoxy (Araldite 2011) is applied manually with a toothpick. 
The glue is distributed in small dots in a quincunx pattern.
Then, the HDI is placed by monitoring its position with the microscope (Figure~\ref{fig:ladder_assembly}).
While the glue is curing, brass weights are put onto the ladder.
To reach minimum material budget, the aimed glue thickness is $\sim \SI{5}{\micro\meter}$.
During prototyping the thickness was measured for each ladder and an average glue thickness of \SI{5\pm4}{\micro\meter} was achieved.

After curing, the electrical connection is established via spTAB on a semi-automatic bonding machine.
Each ladder can be tested with a custom test board with an interface to the flexprints.

\paragraph{Module construction}

The two vertex detector layers are made of half-shell modules.
Modules are made of 4 or 5 ladders, depending if belonging to layer~1~or~2, respectively.
They are supported by a polyetherimide (PEI) endpiece on each side where the ladders are electrically connected to an endpiece flexprint via a Samtec ZA8H interposer\footnote{\url{https://www.samtec.com/products/za8h}}. 

\begin{figure}[tb!]
	\centering
	\includegraphics[width=0.37\textwidth]{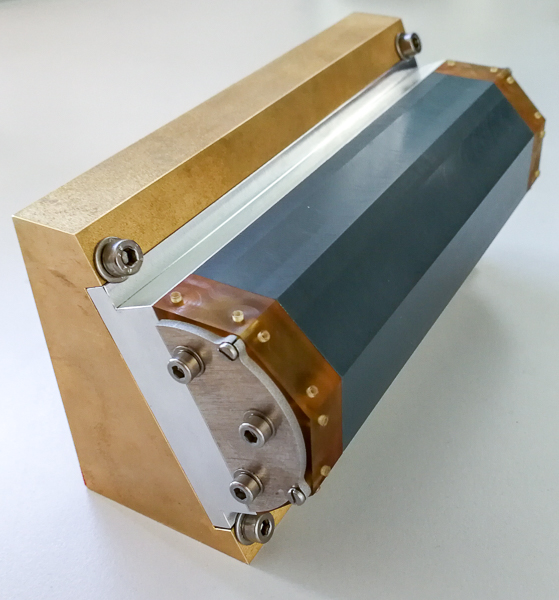}
	\caption{Module assembly tool for layer~2~\cite{tdr}.}
	\label{fig:module_mounting}
\end{figure}

A module is constructed by placing ladders, one by one, on a custom mounting block (Figure~\ref{fig:module_mounting}).
The tool allows to bring any of the facets into a horizontal position by combinations of turning and flipping of the brass block and/or the aluminum slider.

The passive polyimide flaps of the HDIs (Figure~\ref{fig:hdi}) are used to glue neighboring ladders to each other.
This results in a self-supporting structure that can already be mounted in the experiment.
HV, LV and signal lines are connected to the power supplies or front-end boards.

\section{Thermal-mechanical mock-up}
\label{sec:prototypes}

In the past, the general concept of detector construction and cooling was proven~\cite{proceedings_frank}.
It was studied with a thermo-mechanical mock-up using aluminized polyimide tapes equipped with steel plates.

\begin{figure*}[bt]
	\centering
	\begin{subfigure}[c]{.5\textwidth}
		\includegraphics[width=\textwidth]{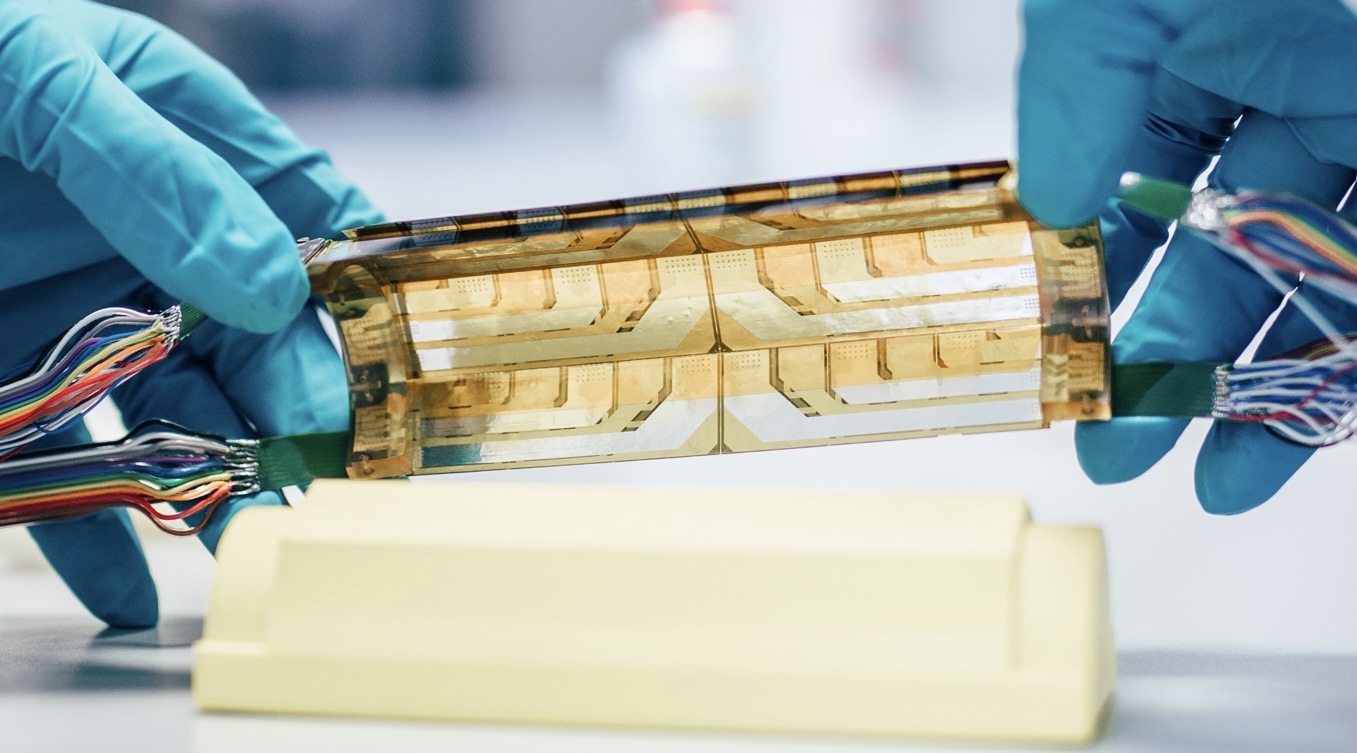}
		\caption{Module of a silicon heater mock-up. View on the inside facing the HDIs.}
		\label{fig:si_heater_mock-up}
	\end{subfigure}
\hspace{1.5em}
	\begin{subfigure}[c]{.26\textwidth}
		\bigskip
		\includegraphics[width=\textwidth]{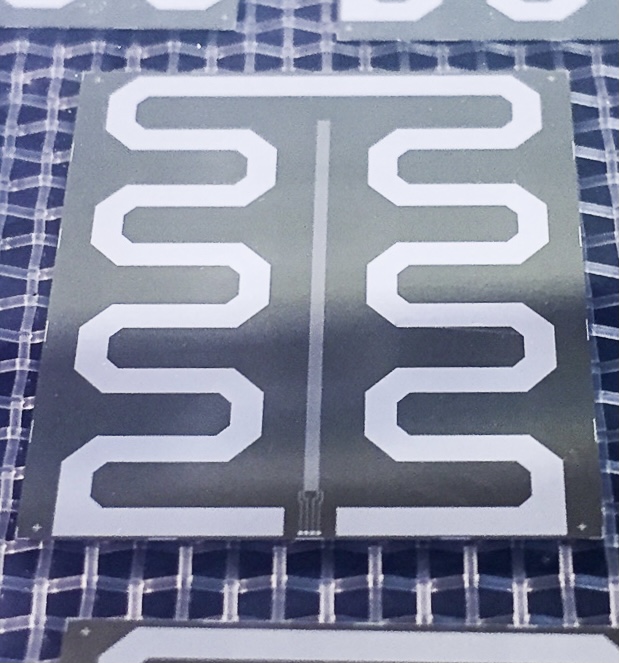}
		\caption{Silicon heater chip with heating loop and resistive thermometer~\cite{tdr}.}
		\label{fig:si_heater}
	\end{subfigure}
	\caption{Silicon heater mock-up and chip}
	\label{fig:silicon_heaters}
\end{figure*}

In 2020, a prototype was constructed that matches all mechanical and geometrical properties of the final detector (Figure~\ref{fig:si_heater_mock-up}).
Main goals of this project were:
First, the verification of the tooling and, second, the verification of the helium cooling concept.
For the mock-up, \SI{50}{\micro\meter} thin silicon heater chips (Figure~\ref{fig:si_heater}) made at MPG Halbleiterlabor in Munich are used, instead of MuPix sensors.
This allows for a controlled heating and temperature measurement via a Al-\SI{1.4}{\kilo\ohm} thermometer.
The HDIs have the same geometry as the final detector.
Each half-ladder (set of three chips) can be powered and thus heated individually. 
The resistive thermometers of each sensor are readout.

The silicon heater ladders are constructed in the same way as the final sensors described in section~\ref{sec:construction}.
All mentioned working steps were successfully verified with this silicon heater mock-up.

\begin{figure}[tb!]
	\centering
	\includegraphics[width=0.8\textwidth]{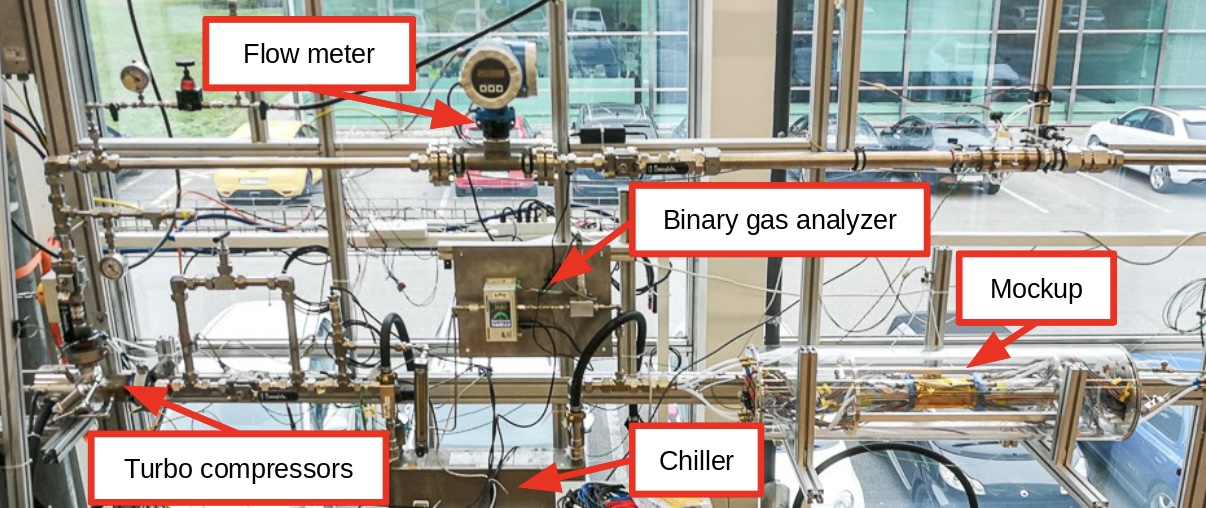}
	\caption{Helium cooling test stand at the Fachhochschule Nordwestschweiz in Windisch, Switzerland. Circulating helium flow established by turbo compressor.}
	\label{fig:setup_brugg}
\end{figure}

The mock-up was mounted in a mounting frame (Figure~\ref{fig:setup_brugg}) and operated at a test stand at the FHNW.
For first measurements, the temperatures of six chips of a ladder in layer~1 were measured since they are expected to get hotter than those from layer~2 which is cooled from both sides.
The heat dissipation was limited to \SI{200}{\milli\watt/\centi\meter^2} since the inlet temperature of the setup was higher than in the final experiment

\begin{figure*}[bt]
	\centering
	\begin{subfigure}[b]{.48\textwidth}
		\includegraphics[width=\textwidth]{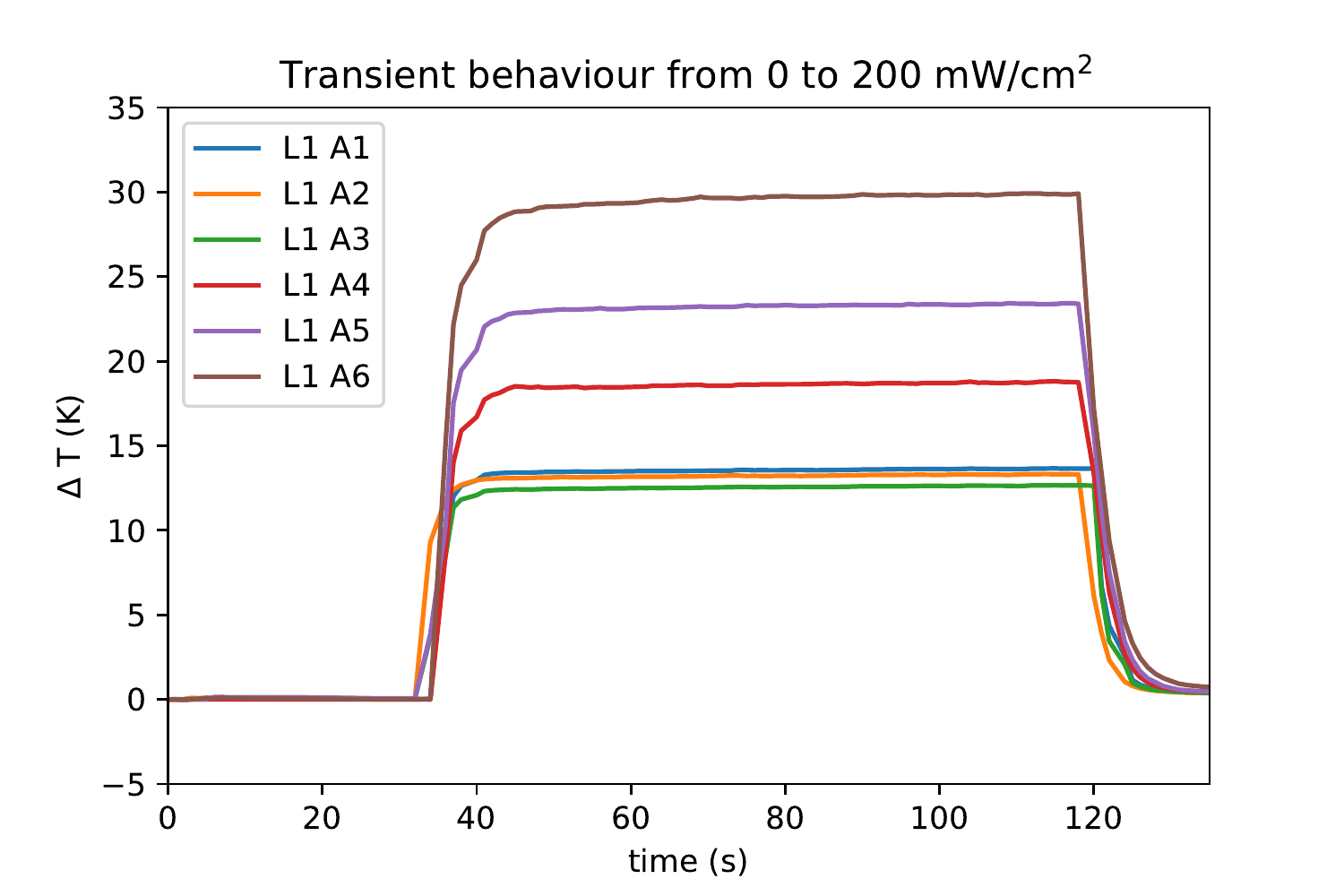}
		\caption{Transient behavior of temperature, stabilization after  a few seconds.}
		\label{fig:transient}
	\end{subfigure}
	\hspace{0.5em}
	\begin{subfigure}[b]{.48\textwidth}
		\bigskip
		\includegraphics[width=\textwidth]{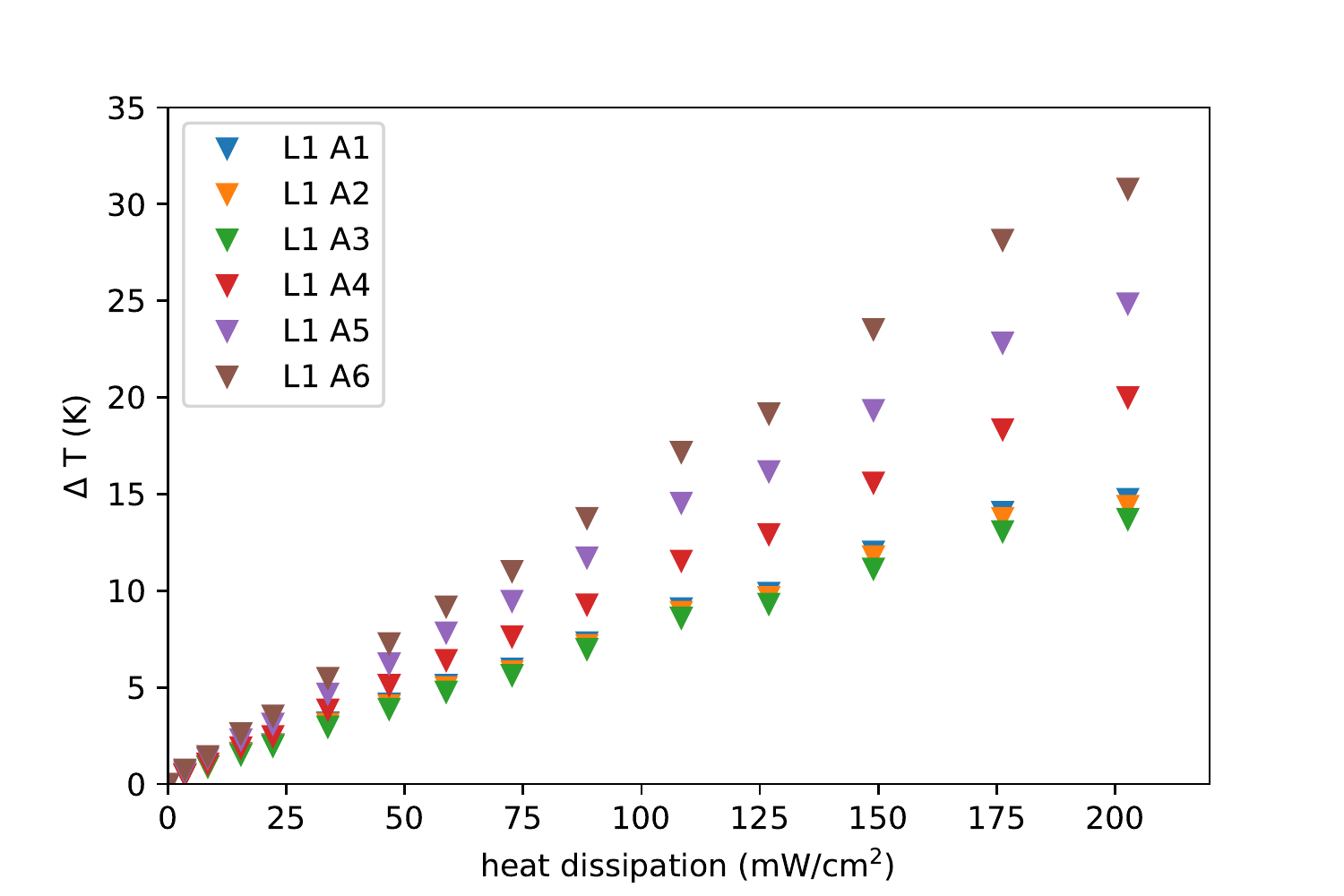}
		\caption{Linear relation between chip temperature and dissipated heat.}
		\label{fig:T-P-relation}
	\end{subfigure}
	\caption{Thermal behavior of six chips on one inner ladder. Chip 1 is located at the helium inlet.}
	\label{fig:silicon_heater_measurement}
\end{figure*}

First preliminary results of the transient behavior as well as the temperature-to-power relation were obtained.
The temperature of the detector for turning on and off at \SI{200}{\milli\watt/\centi\meter^2} is shown in Figure~\ref{fig:transient}.
The ladders reach equilibrium in a few seconds and remain stable. 
No significant fluctuations are observed during operation.
The further the sensor from the inlet, the higher the temperature.

The temperature-to-power relation was measured by slowly increasing the heat dissipation.
Figure~\ref{fig:T-P-relation} shows the expected linear behavior.
The maximum temperature difference reached for a power of  \SI{200}{\milli\watt/\centi\meter^2} is $\Delta T=\SI{30}{\kelvin}$.
One can expect $\Delta T \approx \SI{60}{\kelvin}$ for the conservative scenario of \SI{400}{\milli\watt/\centi\meter^2}.
Considering the foreseen inlet temperature of 0-\SI{4}{\degreeCelsius}, this is well in specs with the maximum allowed temperature of \SI{70}{\degreeCelsius}.

More detailed studies with the mock-up are ongoing at PSI.

\section{Vertex detector prototype}

\begin{figure}[bt!]
	\centering
	\includegraphics[width=0.7\textwidth]{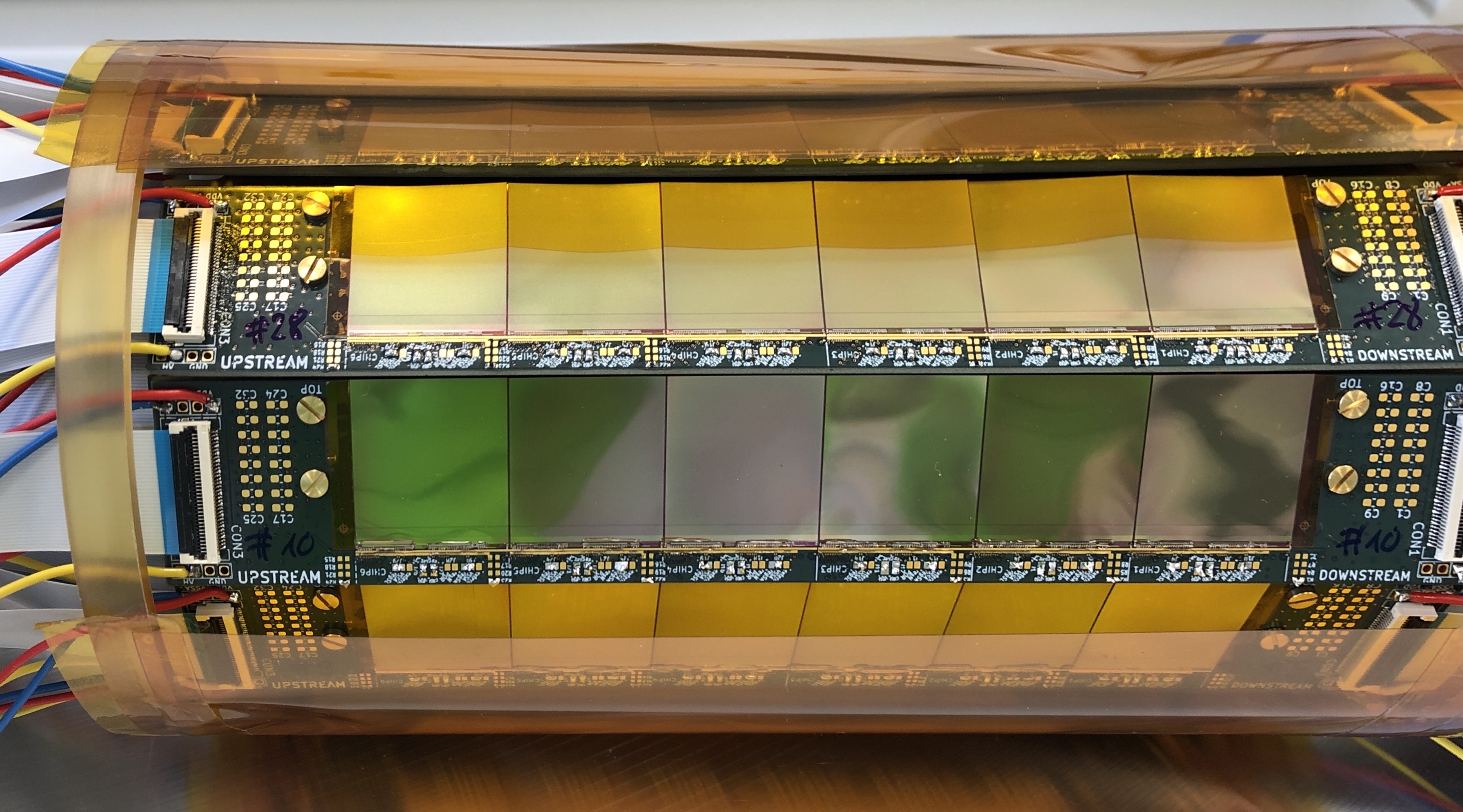}
	\caption{Vertex detector prototype making use of u-shaped PCB ladders equipped with six MuPix10. The outer layer is surrounded by a polyimide foil to contain helium flow.}
	\label{fig:pcb_detector}
\end{figure}

In June and July 2021, the Mu3e collaboration aims to test several detector and DAQ components in a test beam campaign at PSI.
For the pixel group, the goal is to test a two-layered vertex detector equipped with 108 sensors and its readout (Figure~\ref{fig:pcb_detector}).
All sensors have a thickness of \SI{50}{\micro\meter}.

\begin{figure}[bt!]
	\centering
	\includegraphics[width=.7\textwidth]{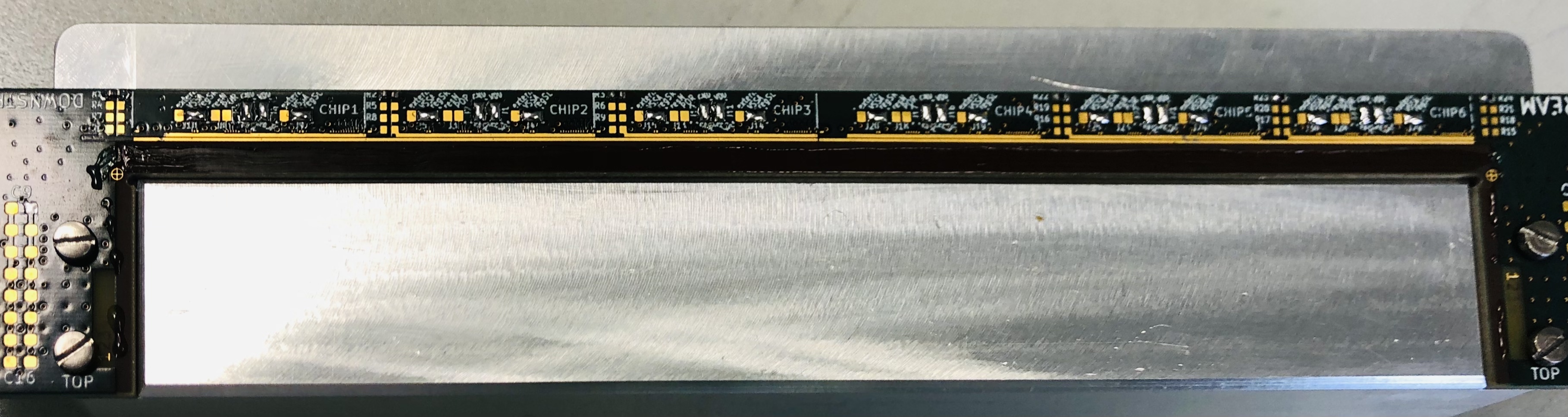}
	\caption{6-chip-PCB on aluminum chuck before gluing of sensors.}
	\label{fig:6-chip-PCB}
\end{figure}

The material budget is not of concern for this test run. 
Thus, u-shaped PCBs (Figure~\ref{fig:6-chip-PCB}) were used to carry the six sensors of a ladder instead of HDIs. 
The sensors are first glued on a polyimide tape which again is glued to the ladder PCB.
The electrical connections are made by conventional wire bonds. 

In the lab, each ladder has been tested with the final readout electronics.
After full assembly, a~yield of  $\sim\SI{82}{\percent}$ configurable sensors with error free data output was reached.
Due to the time schedule imperfect ladders were also accepted for this prototype.

\section{Conclusion}

In the last year, major steps toward production readiness of the Mu3e vertex detector were made.
The final tooling and procedure for the pixel detector construction were developed and tested.
A perfectly matching prototype with non-active silicon chips was built fulfilling the required tolerances.
A prototype of the helium cooling system was successfully operated at the FHNW.

In summer 2021, a vertex detector of simplified geometry will be operated for the first time inside the Mu3e magnet at PSI.

\section*{Acknowledgments}

H. Augustin, A. Meneses Gonz\'{a}lez, D. M. Immig, T. Rudzki, A. Weber, and Benjamin Weinl\"ader acknowledge support by the HighRR research training group [GRK 2058].

The measurements leading to these results have been performed at the Test Beam Facility at DESY Hamburg (Germany), a member of the Helmholtz Association (HGF).

We gratefully thank the technicians and engineers of the workshop at the Physics Institute in Heidelberg who made a tremendous contribution to the realization of this project.

\section*{References}

\bibliography{mybibfile}

\begin{thebibliography}{10}
\expandafter\ifx\csname url\endcsname\relax
  \def\url#1{\texttt{#1}}\fi
\expandafter\ifx\csname urlprefix\endcsname\relax\def\urlprefix{URL }\fi
\expandafter\ifx\csname href\endcsname\relax
  \def\href#1#2{#2} \def\path#1{#1}\fi

\bibitem{mu3e:research_proposal}
A.~{Blondel}, et~al., {Research Proposal for an Experiment to Search for the
  Decay $\mu \rightarrow eee$}, ArXiv e-prints\href
  {http://arxiv.org/abs/1301.6113} {\path{arXiv:1301.6113}}.

\bibitem{sindrum}
U.~Bellgardt, et~al., {Search for the Decay $\mu^+ \rightarrow e^+ e^+ e^-$},
  Nucl.Phys. B299 (1988) 1.
\newblock \href {http://dx.doi.org/10.1016/0550-3213(88)90462-2}
  {\path{doi:10.1016/0550-3213(88)90462-2}}.

\bibitem{hvmaps_ivan}
I.~Peri\'c, {A novel monolithic pixelated particle detector implemented in
  high-voltage CMOS technology}, Nucl. Instrum. Meth. A 582 (2007) 876--885.
\newblock \href {http://dx.doi.org/10.1016/j.nima.2007.07.115}
  {\path{doi:10.1016/j.nima.2007.07.115}}.

\bibitem{proceedings_andre}
A.~Sch\"oning, et~al., {MuPix and ATLASPix -- Architectures and Results}, PoS
  Vertex2019 (2020) 024.
\newblock \href {http://arxiv.org/abs/2002.07253} {\path{arXiv:2002.07253}},
  \href {http://dx.doi.org/10.22323/1.373.0024}
  {\path{doi:10.22323/1.373.0024}}.

\bibitem{tdr}
K.~Arndt, et~al., {Technical design of the phase I Mu3e experiment}, accepted
  for publication in Nucl. Instr. Meth. A (2021).
\newblock \href {http://arxiv.org/abs/2009.11690} {\path{arXiv:2009.11690}}.

\bibitem{bttb9_david}
D.~M. Immig, {The Very Large HV-MAPS Tracking Telescope},
  \url{https://indico.cern.ch/event/945675/contributions/4184873/attachments/2186614/3694678/bttb9_immig.pdf}
  (2021).

\bibitem{ba_frauen}
F.~Frauen, {Characterisation of the time resolution of the MuPix10 pixel
  sensor}, {Bachelor Thesis at the University Heidelberg} (2021).

\bibitem{mupix10}
H.~Augustin, et~al., Mupix10: First results from the final design, submitted to
  JPS (2020).
\newblock \href {http://arxiv.org/abs/2012.05868} {\path{arXiv:2012.05868}}.

\bibitem{proceedings_frank}
F.~Meier~Aeschbacher, et~al., {Mechanics, readout and cooling systems of the
  Mu3e experiment}, PoS Vertex2019 (2020) 041.
\newblock \href {http://arxiv.org/abs/2003.11077} {\path{arXiv:2003.11077}},
  \href {http://dx.doi.org/10.22323/1.373.0041}
  {\path{doi:10.22323/1.373.0041}}.

\bibitem{LTU}
{LTU, LED Technologies of Ukraine}, \url{http://ltu.ua/en/index/}.

\end{thebibliography}

\end{document}